# Deformation of the Bodies by the Result of Length Contraction: A new Approach to the Lorentz Contraction

Bayram Akarsu, Ph.D
Erciyes University
Kayseri/ Turkiye

Abstract

It has been more than a century since first Lorentz and later Einstein explored relativistic events and still important consequences of that remains unclear to everybody.
The present study extensively focus on Lorentz (Length) contraction phenomenon and takes a different approaches to explain how the shrinking of the body appears if exists. We utilized two postulates of Special Relativity that emphasizes constancy of speed of light and same physical laws in any inertial frame. Finally, we propose Lorentz contraction exists in different form through the shape of the accelerated body. The deformation of its shape depends on the part where it is measured.

I. Introduction

In his Dialogue Concerning the Two Chief World Systems, Galileo Galilei, in 1632, described Relativity with a ship model which travels at constant speed, without rocking, on a smooth sea; any observer doing experiments below the deck would not be able to tell whether the ship was moving or stationary. This idea later became Einstein's one of his postulates in his new relativity about three centuries later. However, this explanation of his results didn't guide him full understanding of relativistic picture because one side of his idea was missing regarding speed of light. That latter idea was discovered by Michelson-Morley experiment in the mid $19^{th}$ century that shed light on fully understanding relativity concepts.

Several new papers and articles were published that discussed the consequences of Einstein's relativity but still some points remain unclear to the physicists and more importantly to physics educators. The study of length contraction and time dilation goes back to 1950s with the paper of Terrell (1986) paper which discusses relativistic consequences of the invisibility of Lorentz contraction. The author showed that anybody could deduce the formulas and claimed that even though one can use methods of various measuring techniques (e.g. stereoscopic photography), length contraction cannot be pictured.

Boas (1961) strongly believed that the relativistic contraction be seen as curved not shrinking. Boas focused on famous Gamow's bicyclist example to apply her ideas to show that the shape of the picture is curved rather than short. Based on the outcomes of previous research ideas, the current paper supports Boas' ideas from different perspectives. We strongly believe that shape of the moving body appears curved from both ends and we will show that by using time dilation ideas and constancy of light.

Later, Rindler (1961) discussed the deformation of the accelerated body with an example of a man walking very fast and tries to get over a deep and wide well which certainly results in unsuccessful trials. He calls it a paradox but there is no paradox because the moving body is not in the same inertial frames as the grid. Grids are present in the observer's frames of references. However, I agree with other results which reveal that the warping of the shape of the moving body starts with the front of the body pointed towards to the front.

II. Claim

We strongly believe that Lorentz contraction exists but we cannot suggest it can be measured or photographed in the same sense because the apparatus currently used is not enough to reveal it really occurs in natural environment. We propose a new approach to the change of shape of the accelerated body. We make use of train example. Consider we have a wagon in a moving train with a relativistic velocity and we have two mirrors place diagonally at the left lower and upper front of the wagon in figure 1. We generate a light signal from rear end of the wagon and it is reflected back from the front mirror. An observer outside next to train station observes light signal while being at rest (figure 2). We compare the measurement of time intervals by train clock and observer's time.

Figure 1. Bouncing light rays from the diagonal mirrors as seen by a person in the wagon

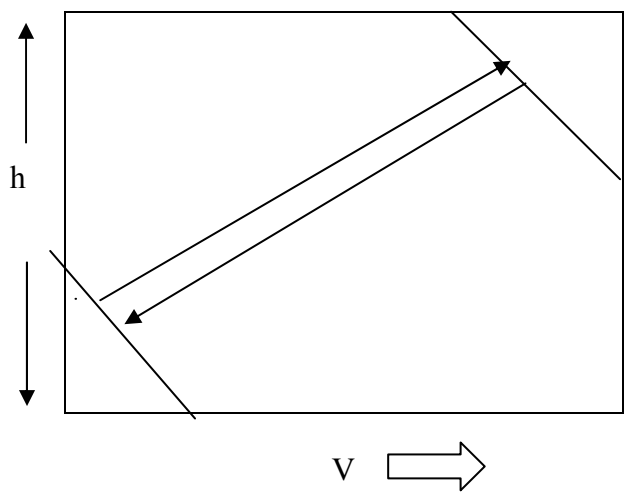

Total distance of outgoing light ray and height of the wagon are equal to:

ℓ = v × t (t is the proper time inside the train)   (1)

Let τ be time according to outside observer's clock so related time differences between two persons, (time dilation equation)

τ = γ × t   (2)

Also, we can easily write down the height of the wagon by using the total travel times of light rays and the speed of light,

h = c × t   (3)

The equations 1-3 belong to the figure one and based on the observer inside the wagon. Now, we can look at the equations constructed according to the out side observer.

Figure 2. Bouncing light rays from the diagonal mirrors in the wagon as seen by outside observer

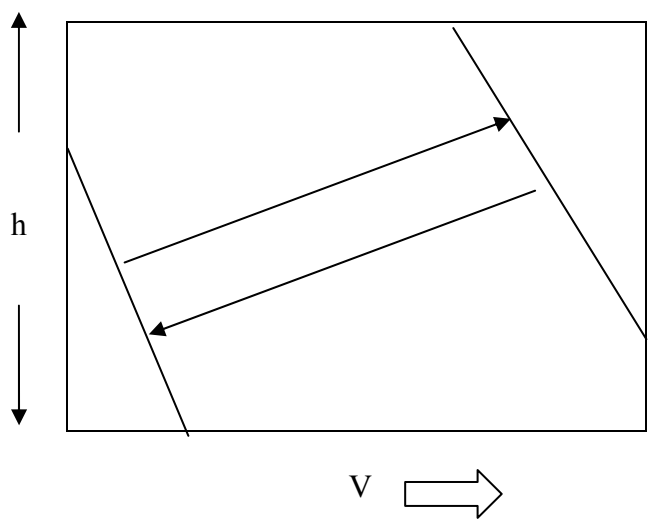

Total distance of outgoing light ray according to the observer at the train station is:

$\ell' = c \times \tau$ (4)

Let us suppose the speed of the wagon is close to the speed of light so that angle between light rays and horizontal axis of the wagon is zero. We can easily construct a right angle triangle and by using equations 1, 3, and 4 (Pythagoras theorem)

$(\ell)^2 + (\ell')^2 = (h)^2$

That gives us the famous length contraction formula,

$c^2 t^2 - c^2 \gamma^2 t^2 = \ell^2$

$\ell' = \ell \times \gamma$ ($\ell'$: length of the wagon measured by the observer at the station)     (5)

Now that we proved the deformation and shrinking of the wagon with the usage of time dilation we can look at the corners of the wagon however their shapes change during the motion.

We can construct a different, than the prior case, right angle triangle. This time the wagon travels at 0.5c (half the speed of light). Based on that, the light rays at both cases (inside observer and outside observer) do not create a right angle triangle but if we include the length of the wagon it construct two triangles as Figure 3,

Figure 3. Contruction of two triangles at the speed of half c

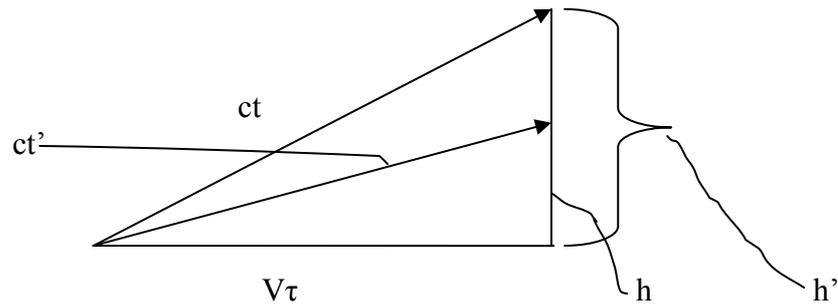

The equations are constructed by two right angle triangles,

$(ct)^2 - (V\tau)^2 = (h)^2$ (6)

$(c\tau)^2 - (V\tau)^2 = (h')^2$ (7)

By eliminating V terms first and time terms next in equations 6 and 7 with $\tau = \gamma t$, we can easily deduce the following equations,

$(h')^2 + (h)^2 = (ct)^2 - (c\tau)^2$

$(h')^2 - (h)^2 = 2(V\tau)^2 - (V\tau)^2$

Finally, we can deduce the deformation formulas as,

$(h')^2 = (ct)^2 - (V\tau)^2$ (8)

$h = ct$ (9)

Equations 8 and 9 proves the shape changing of the wagon and give an approximation of the rate of change and it depends on where it is measured at the wagon. Finally, at the speeds close to the speed of light, the rectangular train wagon will look like parallelogram and oval,

Figure 4. Formation shape of the wagon at the speeds close to the speed of light

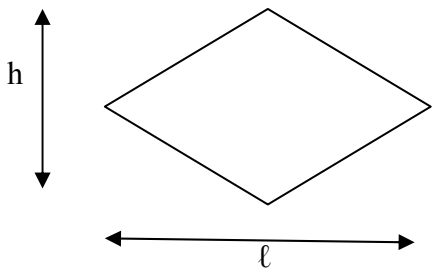

III. Conclusion

In this article, only one of the SR postulates, the constancy of the speed of light, was used to implement a new perspective of the length contraction phenomenon according to the Lorentz formulas. The method used to explain the deformation of the objects because of their high speeds is a very simple example and can be found in several elementary SR textbooks to derive the time dilation equation. However, we used and generalized it by looking at the light rays with various velocities as they reach the speed of light. This new more generalized method can be used as more advanced approach in relativity books to illustrate the deformation degrees. We don't claim that this deformation is visible or can be photographed. We will investigate that in a separate and further study.

In conclusion, we claim that the angle of the light rays (both incoming and reflected) will create smaller angles with the horizontal length of the wagon because of the first postulate of Einstein's SR, constancy of the speed of light, and it is obvious that this angle will get smaller as the wagon speeds up and it will reach to cut-off value 0 as the wagon travels at the speed of light. Hence, we expect major deformation from both ends of the wagon shaping it like a two rockets connected to each other at rear ends.

Our assertion creates a new approach to length contraction phenomena but it show similarities to the previous research findings by Rindler (1961). The only difference between our expectations and his is we expect both ends of the wagon will shrink at some level but the same rate and also the length of the wagon will reduce in size respectively.